\documentclass[superscriptaddress,aps,prx,twocolumn,showpacs,nofootinbib, longbibliography, notitlepage,floatfix]{revtex4-2}
\usepackage{etex}
\usepackage{amsmath,amssymb,amsthm}
\usepackage[colorlinks=true,citecolor=blue,urlcolor=blue]{hyperref}
\usepackage[pdftex]{graphicx}
\usepackage{times,txfonts}
\usepackage{braket}
\usepackage{color}
\usepackage{natbib}
\usepackage{bm}
\usepackage{soul}
\usepackage{amsmath,blkarray}
\usepackage{mathtools}
\usepackage{latexsym}
\usepackage{tabularx, booktabs}
\usepackage{graphics,epstopdf}
\usepackage{graphicx}
\usepackage{float}
\usepackage{amsfonts}
\usepackage{tikz}
\usetikzlibrary{quantikz}
\usepackage{color,soul}

\newcommand{\be}{\begin{equation}}
\newcommand{\ee}{\end{equation}}
\newcommand{\ba}{\begin{eqnarray}}
\newcommand{\ea}{\end{eqnarray}}

\usepackage{multirow}
\usepackage{appendix}
\usepackage{url}
\usepackage{listings}

\usepackage{comment}

\begin{document}
\title{Efficient DCQO Algorithm within the Impulse Regime for Portfolio Optimization}

\author{Alejandro Gomez Cadavid}
\affiliation{Kipu Quantum, Greifswalderstrasse 226, 10405 Berlin, Germany}

\author{Iraitz Montalban}
\affiliation{Kipu Quantum, Greifswalderstrasse 226, 10405 Berlin, Germany}
\affiliation{Department of Physics, University of the Basque Country UPV/EHU, Barrio Sarriena, s/n, 48940 Leioa, Biscay, Spain}

\author{Archismita Dalal}
\affiliation{Kipu Quantum, Greifswalderstrasse 226, 10405 Berlin, Germany}

\author{Enrique Solano}
\email{enrique.solano@kipu-quantum.com}
\affiliation{Kipu Quantum, Greifswalderstrasse 226, 10405 Berlin, Germany}

\author{Narendra N. Hegade}
\email{narendra.hegade@kipu-quantum.com}
\affiliation{Kipu Quantum, Greifswalderstrasse 226, 10405 Berlin, Germany}

\begin{abstract}
We propose a faster digital quantum algorithm for portfolio optimization using the digitized-counterdiabatic quantum optimization (DCQO) paradigm in the impulse regime, that is, where the counterdiabatic terms are dominant.  Our approach notably reduces the circuit depth requirement of the algorithm and enhances the solution accuracy, making it suitable for current quantum processors. We apply this protocol to a real-case scenario of portfolio optimization with 20 assets, using purely quantum and hybrid classical-quantum paradigms. We experimentally demonstrate the advantages of our protocol using up to 20 qubits on an IonQ trapped-ion quantum computer. By benchmarking our method against the standard quantum approximate optimization algorithm and finite-time digitized-adiabatic algorithms, we obtain a significant reduction in the circuit depth by factors of 2.5 to 40, while minimizing the dependence on the classical optimization subroutine. Besides portfolio optimization, the proposed method is applicable to a large class of combinatorial optimization problems.
\end{abstract}

\maketitle
\section{Introduction}

Quantum computing aims at challenging the status quo of classical computation for hard-to-solve problems, as it was first demonstrated in Shor's quantum algorithm for prime factorization \cite{shor1994algorithms}. However, the hardware requirements raise valid concerns on scalability issues and the effective translation from theoretical to experimental realizations \cite{gidney2021factor}. In this sense, when using noisy-intermediate scale quantum (NISQ) processors, it is unavoidable to consider specific-purpose implementations of the targeted problem \cite{bharti2022noisy}. Our NISQ era needs pragmatic solutions via the codesigned encoding of compressed algorithms in available hardware.

Combinatorial optimization problems relate to many challenges faced by companies in various industries, such as logistics, finance, and chemistry, among others \cite{orus2019quantum, mcardle2020quantum}. This explains the wide attention and methods developed, where the quantum approximate optimization algorithm (QAOA) \cite{farhi2014quantum} is one of the most adopted variational protocols \cite{zhou2020quantum}. QAOA is one of the hybrid classical-quantum approaches known as variational quantum algorithms (VQA), and it requires classical training to fit an ansatz or parametrized quantum circuit (PQC) composed of discrete gates. Similarly to Shor's algorithm, it does suffer from discrepancies between theoretical proposals \cite{farhi2022quantum, boulebnane2022solving} and hardware implementations \cite{harrigan2021quantum}. Given its canonical form, it produces relatively deep quantum circuits that are known to be impacted by hardware noise \cite{wang2021noise}. There is extensive literature on how PQCs can be trained and modified in order to circumvent those issues \cite{streif2020training, larocca2022diagnosing, sack2023large}. 

The prospect of achieving a quantum advantage in finance use cases is being extensively explored~\cite{orus2019quantum, herman2023quantum}. In particular, the portfolio optimization problem is a non-trivial and interesting use case for NISQ testbeds with different strategies~\cite{egger2021warm, brandhofer2022benchmarking, HSC+23}. Both analog~(quantum annealing) and digital~(QAOA) algorithms are commonly used for solving this constrained optimization problem. The state-of-the-art implementation of QAOA for portfolio optimization shows the convenience of using a constraint-preserving mixer Hamiltonian, which strongly improves the approximation ratio~\cite{HSC+23}.

In this paper, we present an efficient quantum computing solution for portfolio optimization using a fast purely-quantum digitized-counterdiabatic quantum optimization (DCQO) protocol, as well as its hybrid counterpart h-DCQO. Our results show an important improvement on previous results \cite{hegade2021shortcuts, vcepaite2023counterdiabatic, chandarana2022protein, chandarana2022digitized, vizzuso2023convergence, ding2023molecular}, as we explain in detail along this article. We rely on the concept of the impulse regime, which corresponds to the fast non-adiabatic evolution where the counterdiabatic~(CD) terms are dominant. Our method reduces the circuit depth and increases the solution accuracy with respect to finite-time digitized-adiabatic algorithms and the conventional QAOA. In particular, we successfully test our protocol for a 20-asset portfolio optimization problem on IonQ quantum processors.

The paper is organized as follows. 
In \S\ref{sec:prelim}, we briefly describe the problem of portfolio optimization, the techniques of adiabatic quantum optimization, CD protocols and their digitized versions. We also discuss QAOA and its extension using CD protocols.
Next in \S\ref{sec:approach}, we elaborate on our proposed algorithms of DCQO and h-DCQO and introduce our modified definition for approximation ratio.
Then we present our results, from both simulation and experiment, and discuss them in \S\ref{sec:results}.
Finally, we conclude with an outlook in~\S\ref{sec:conclusion}.

\section{Portfolio optimization and DCQO}
\label{sec:prelim}
\subsection{Portfolio optimization}

Within the financial services and banking industry that emerged with Markowitz's seminal paper in 1952 \cite{markowitz1952}, portfolio optimization occupies a central role. It describes a set of $n$ assets $x_i \in \{X\}_n$ from which a subset needs to be picked. The goal is to maximize the revenue while minimizing the risk at $t$ future time steps. Each asset~$i$ has an associated revenue forecast at a given time period ($e_i$) and the covariance between assets ($c_{ij}$) sets the risk amount in terms of diversification.

For the sake of simplicity, we will focus on a single-time step modality of this problem with Boolean asset investment. Let us consider then, a single time step $t$ and a budget $B$, which defines the number of assets that can be selected. The budget sets a constraint to the problem, which is described by the summation cost $x_i$ of investing on each individual asset $i$. Ideally, the solution should not surpass this budget condition ($\sum_i x_i \leq B$). Therefore, the problem is posed in the first stage as follows
\begin{equation}\label{eq:canonical}
\begin{array}{rrclcl}
\displaystyle \max_{x_{i}\in{\{0,1\}}} & \multicolumn{3}{l}{\sum_{i=1}^{n} x_{i}e_i -\;\theta \sum_{i,j=1}^{n}x_{i}x_{j}c_{ij}}\\
\textrm{s.t.} & \sum_{i=1}^{n}x_{i} \leq B \, . \\
\end{array}
\end{equation}
Here,  $x_{i}\in{\{0,1\}}$ is the mask associated with the selection of our set of assets, while $\theta$ is a Lagrange multiplier modulating the amount of risk we would like to assume in our investments.

Single-period discrete portfolio optimization has been shown to be NP-complete regardless of the risk \cite{kellerer2000selecting}. Several classical solutions have been proposed for this problem. However, quantum computing may produce faster solutions due to the connection between spin glasses and Markowitz portfolio optimization \cite{herman2023quantum}.

To convert our canonical formulation of Eq.~(\ref{eq:canonical}) to a Hamiltonian formulation, the problem is reformulated as follows
\begin{equation}\label{eq:portfolio}
\max_{x_{i}\in{\{0,1\}}} \;\; \theta_{1}\sum_{i=1}^{n} x_{i}e_i \\
\;\;  -\;\theta_{2}\sum_{i,j=1}^{n}x_{i}x_{j}c_{ij}  \\
 -\;\theta_{3}\left(\sum_{i=1}^{n}x_{i}-B\right)^{2} .
\end{equation}
Here, the budget constraint has been integrated to the cost function and $\theta_{1,2,3}$ are Lagrange multipliers, which decide the relevance of each term. In this way, one can modulate not only the risk but also the option of surpassing the budget if necessary, being able to simulate all potential scenarios for a decision to be made regarding the portfolio.
The objective function in Eq.~(\ref{eq:portfolio}) can then be easily translated into an Ising formulation that simplifies the encoding~\cite{elsokkary2017financial, parizy2022cardinality} of the problem in future steps.

\subsection{Adiabatic quantum optimization and CD protocols}

Computing the optimal solution for Eq.~(\ref{eq:portfolio}) requires to find the ground state of an Ising spin-glass Hamiltonian $H_\text{f}$, which we call the problem Hamiltonian,
\begin{equation}\label{eq:hamiltonian}
H_\text{f} = \sum_{i=1}^{n} h_i Z_i + \sum_{i<j} J_{ij} Z_i Z_j .
\end{equation}
Its ground state encodes the binary string, or strings in case of degeneracies, as the mask of assets to be chosen. In this sense, we use the adiabatic theorem to find this ground state \cite{lopez2015generalized, rosenberg2015solving}. Here, a quantum system slowly evolves from the ground state of an initial Hamiltonian, chosen in this case as $H_\text{i} = -\sum_{i=1}^n X_i$ with ground state $\left( \frac{\ket{0} + \ket{1}}{\sqrt{2}}\right )^{\otimes n}$, towards the ground state of the problem Hamiltonian $H_\text{f}$. This evolution is governed by the Hamiltonian 
\begin{eqnarray}
\label{eq:adiabatic}
    H_\text{ad}(t) = \left[1 - \lambda(t) \right ]H_\text{i} + \lambda(t) H_\text{f},
\end{eqnarray}
where $\lambda(t)$ is a scheduling function satisfying the conditions $\lambda(0)=0$ and $\lambda(T)=1$, with $T$ the evolution time. In this manner, it is ensured that the system goes from the ground state of $H_\text{i}$ to the ground state of $H_\text{f}$. As a consequence, the measured state at the end of the process will yield the solution to the optimization problem. Along this article, we use the following scheduling function $\lambda(t) = \sin^2\left[ \frac{\pi}{2} \sin^2(\frac{\pi t} {2T} )\right ]$, such that $\dot{\lambda}(0) = \dot{\lambda}(T) = 0$.

The process described in Eq.~(\ref{eq:adiabatic}) can be encoded in quantum annealers, which are analog devices that have been extensively used in order to tackle various optimization problems \cite{yarkoni2022quantum}. By nature, quantum annealing is a slow process, therefore affected by decoherence, which leads to an accumulation of errors. To mitigate this issue, one might resort to rapid non-adiabatic evolution. However, performing fast evolution of Eq. (\ref{eq:adiabatic}) causes excitations between eigenstates, compromising the quality of the final solution.

Counterdiabatic driving is a method that helps to suppress non-adiabatic transitions due to rapid evolution~\cite{demirplak2003adiabatic, berry2009transitionless, del2013shortcuts}. This driving is introduced as an extension of Eq.~(\ref{eq:adiabatic}) and can be written in the form
\begin{equation}\label{eq:counterdiabatic}
    H(t) = H_\text{ad}(t) + \dot{\lambda}(t) A_{\lambda} .
\end{equation}
Here, $A_{\lambda}$ is known as the adiabatic gauge potential, responsible for suppressing the non-adiabatic excitations. The calculation of the gauge potential is computationally expensive since it requires knowledge of the instantaneous eigenspectrum of $H_\text{ad}$~\cite{berry2009transitionless}. However, taking the proposal from Ref.~\cite{sels2017minimizing, claeys2019floquet, hatomura2021controlling, takahashi2023shortcuts} one can construct approximate CD protocols that can be easily obtained and experimentally realized. For instance, the gauge potential from Eq.~(\ref{eq:counterdiabatic}) can be approximated by a series of Nested Commutators (NC), i.e.
\begin{equation}\label{eq:commutator}
    A_{\lambda}^{(l)} = i\sum_{k=1}^l \alpha_k(t) \underbrace{[ H_\text{ad}, [ H_\text{ad},\dots [ H_\text{ad},}_{2k-1} \partial_\lambda H_\text{ad} ] ] ],
\end{equation}
where $\alpha(t)$ is a CD coefficient to be found and $l$ sets the order of approximation of the gauge potential. For example, for the Hamiltonians $H_\text{i}$ and $H_\text{f}$ mentioned before, the first order ($l=1$) approximation of the gauge potential is $A_\lambda^{(1)} = - 2 \alpha_1 \left[ \sum_i h_i Y_i + \sum_{i<j} J_{i j}\left(Y_i Z_j+Z_i Y_j\right) \right]$, where
\begin{widetext}
\begin{align}
\alpha_1 = -\frac{1}{4} \frac{ \sum_i h_i^2+ \sum_{i<j} J_{i j}^2 }{(1-\lambda)^2\left(\sum_i h_i^2+4 \sum_{i \neq j} J_{i j}^2\right) + \lambda^2\left[\sum_i h_i^4+\sum_{i \neq j} J_{i j}^4+6 \sum_{i \neq j} h_i^2 J_{i j}^2 +6 \sum_{i<j<k}\left(J_{i j}^2 J_{i k}^2+J_{i j}^2 J_{j k}^2+J_{i k}^2 J_{j k}^2\right)\right] } \, .
\end{align}
\end{widetext}
The detailed calculation of the analytical expression of $\alpha_1$ is given in Appendix~\ref{appendix:a}.

To efficiently implement the counterdiabatic protocols from Eqs. \eqref{eq:counterdiabatic} and \eqref{eq:commutator}, we adopt a digital approach, discretizing the total evolution time \(T\) into \(N\) small steps, each of duration \(dt\). During each brief time interval, the Hamiltonian can be viewed as time-independent. This method allows for the realization of arbitrary counterdiabatic terms, which are not achievable with current analog quantum computers. By employing first-order Trotterization, the continuous-time dynamics can be approximated as
\begin{equation}
U_{\text {digital }}=\prod_{m=1}^N \prod_{k=1}^L \exp \left[-i \, dt \, c_{k}(m dt) \, H_{k}\right].
\end{equation}
Here, $H_k$ stands for each k-local Pauli operator corresponding to the complete Hamiltonian expressed as $H(t) = \sum_{k} c_{k} (t)H_k$. Each matrix exponential term in the above expression can be easily decomposed into a set of quantum gates. The accuracy of this digitization depends on the step size $dt$, with smaller values generally offering improved fidelity at the expense of requiring a larger number of gate sequences. The accuracy and robustness of this digitization method, for both digitized adiabatic evolution and digitized counterdiabatic evolution, have been examined in recent studies \cite{kovalsky2023self}. Furthermore, more efficient methods known as commutator product formulas have been introduced recently to realize digitized-counterdiabatic evolution \cite{chen2022efficient}.

\subsection{Hybrid approaches}
Purely quantum approaches may often require long circuit depths coming from long evolution times, thus exceeding the coherence time of the qubits. This challenge led to the development of hybrid classical-quantum approaches, harnessing the power of classical and quantum devices while looking for optimal solutions using shorter quantum circuits. This will greatly reduce the cumulative effect of noise on existing hardware.

In Ref.~\cite{farhi2014quantum}, the authors proposed the QAOA algorithm, which is inspired by an adiabatic quantum computing procedure that approximates the solution to a combinatorial problem by a set of discrete steps. This is done by combining a problem Hamiltonian and a mixer Hamiltonian, equivalent to $H_\text{f}$ and $H_\text{i}$ in Eq.~(\ref{eq:adiabatic}), respectively. This algorithm has become in recent years the standard technique for most combinatorial problems tackled on near-term quantum devices, besides quantum annealers~\cite{yarkoni2022quantum}.

The QAOA algorithm starts by initializing the qubits in the ground state of $H_\text{i}$, which is a superposition of all possible $2^n$ states, i.e. $|+\rangle^{\otimes n}$. Then, following alternating layers of the unitary associated to the problem Hamiltonian $U_f(\gamma) = e^{-i\gamma H_\text{f}}$, parameterized by $\gamma$, and a mixer unitary $U_i(\beta) = e^{-i\beta H_\text{i}}$, parameterized by $\beta$, the algorithm produces the following state 
\begin{equation}\label{eq:alternating}
|\bm{\gamma, \beta} \rangle = U_i(\beta_p)U_f(\gamma_p) \dots U_i(\beta_1)U_f(\gamma_1)|+\rangle.
\end{equation}
In the same fashion, the hybrid counterpart of the evolution associated with Eq.~(\ref{eq:counterdiabatic}) introduces a new block to the alternating layers that account for the counterdiabatic term $U_c(\alpha)$, which is then parametrized by $\alpha$,
\begin{align}\label{eq:alternating_cd_qaoa}
|\bm{\alpha, \beta, \gamma} \rangle = U_c(\alpha_p)&U_i(\beta_p)U_f(\gamma_p)\notag \dots \\ &U_c(\alpha_1)U_i(\beta_1)U_f(\gamma_1)|+\rangle. 
\end{align}
The extension of QAOA using CD protocols, known as DC-QAOA, was introduced in Refs.~\cite{chandarana2022digitized, wurtz2022counterdiabaticity, yao2021reinforcement}, which has been applied to various problems, e.g. molecular docking~\cite{ding2023molecular}. Recent research also highlights its improved performance when using higher-order CD terms~\cite{vizzuso2023convergence}. In the next section, we will explain how the DC-QAOA method can be further refined by building the variational ansatz using only the CD term. This approach significantly reduces the circuit depth, resulting in better performance, particularly when run on real hardware.

\section{Methodology}
\label{sec:approach}
In this section, we elaborate on the design and analysis of counterdiabatic protocols in the impulse regime for the problem of portfolio optimization. We will describe the two implementations of digitized-counterdiabatic quantum optimization; one being purely quantum and the other being a quantum-classical hybrid algorithm.
Finally, we explain how we model the portfolio optimization problem and propose an efficient quantum algorithm.
\begin{figure*}[ht!]
    \includegraphics[width=0.95\linewidth]{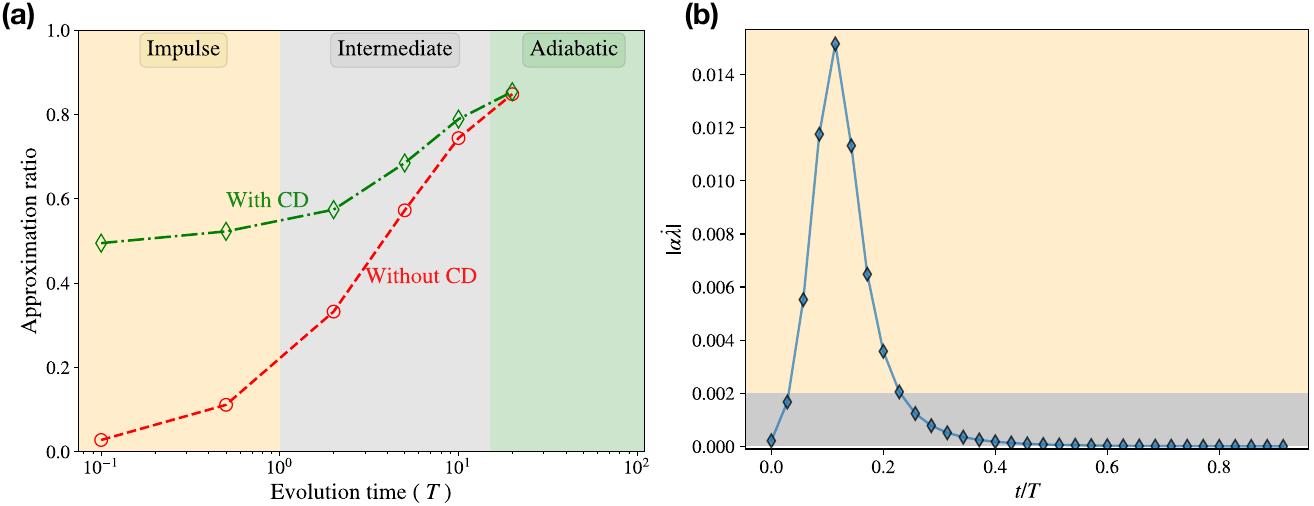}
      \caption{For the purely quantum case, we show the simulated (a) three regimes of evolution under adiabatic Hamiltonian and adiabatic Hamiltonian + CD assistance. The impulse regime (yellow), where CD is dominant over Adiabatic, the intermediate regime (gray), where CD starts to lose dominance until the adiabatic regime (green), where CD is no longer needed nor offers any advantage. (b) Magnitude of $\alpha\dot{\lambda}$, which measures the dominance of the CD term. An estimation of the region in which CD has the most important contribution is highlighted in yellow whereas the one that can be ignored is in grey. The selection of this region would mostly depend on the resource limitations. These results were obtained using a selected $20$ qubit portfolio optimization instance.}
  \label{fig:cdcontrib}
\end{figure*}

\subsection{DCQO in the impulse regime}
\begin{figure*}
  \includegraphics[width=0.9 \textwidth]{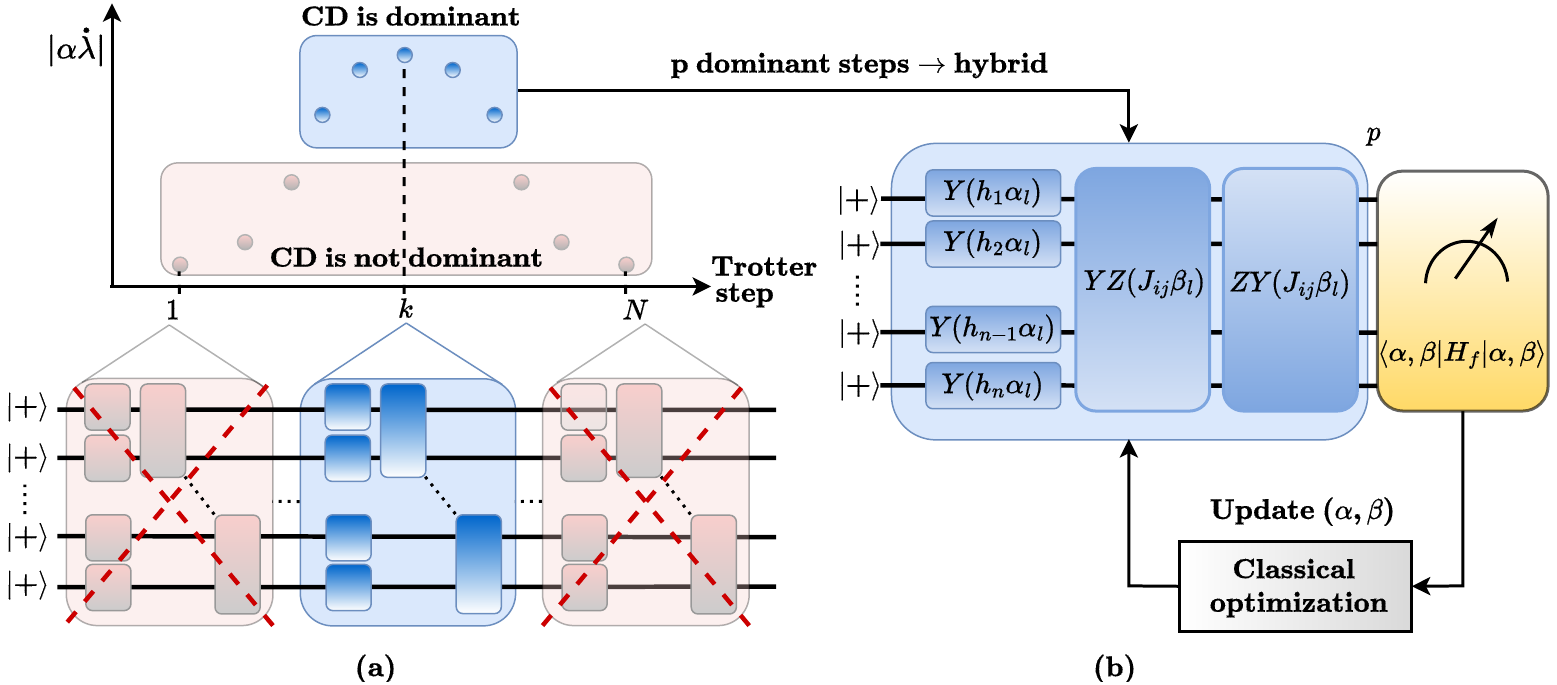}
  \caption{(a) Purely quantum DCQO circuit. In the upper part, a qualitative behavior of $|\alpha \dot{\lambda}|$ is presented, which quantifies the dominance of the CD term. The region where it is dominant (light blue) and where it is not (light red) is shown and mapped to a quantum circuit in the bottom part. Precisely, the first, $k$-th, and last ($N$-th) step is shown. The circuit blocks coming from the first and last steps are not applied since CD is not dominant, whereas the one coming from $k$-th step is applied. The one-qubit blocks represent $R_y$ rotations and the two-qubit ones represent $R_{zy}$ followed by $R_{yz}$. (b) Hybrid DCQO ansatz, which has $p$ layers that mimic $p$ points from the dominant region of the CD term and assigns two parameters per layer, one for single qubit rotations ($\alpha$) and another one for two-qubit rotations ($\beta$). Afterward, the usual hybrid procedure is applied to optimize the cost function $\langle \bm{\alpha,\beta} | H_\text{f} | \bm{\alpha,\beta} \rangle$. Both circuits are initialized by the equal superposition of states, continued by the $Y+YZ+ZY$ operator blocks obtained by the digitization of the CD term for the Ising type of Hamiltonians.}
  \label{fig:hdcqc_ansatz}
\end{figure*}

We analyze the evolution of the CD-assisted Hamiltonian~\eqref{eq:counterdiabatic} in three regimes, namely impulse, intermediate, and adiabatic regimes~\cite{FH_counterdiabatic}. These regimes are characterized by the total evolution time, with short times lying in the impulse regime and long times falling in the adiabatic regime. 
In the impulse regime, the rate of change of the Hamiltonian is rapid, making the CD term more dominant than $H_\text{ad}$ as opposed to the adiabatic regime, where the CD term is inconsequential; see Fig.~\ref{fig:cdcontrib}(a).
Additionally, this observation is also justified by the fact that the two terms in Eq.~(\ref{eq:counterdiabatic}), namely $H_\text{ad}(t)$ and $A_\lambda$, shift their relevance depending on their coefficients.
Thus in the impulse regime, where $|\lambda| << |\dot{\lambda}|$, the CD term~($A_\lambda$) dominates and the evolution is described by
\begin{equation}
\label{eq:impulse}
    H(t) \approx \dot\lambda(t)A_\lambda.
\end{equation}
In our proposal, we work in the impulse regime, see Fig.~\ref{fig:cdcontrib}(b). In addition to that, out of $N=T/dt$ Trotter steps, we only consider $p = N-q$ steps where the CD terms have significant contributions. The remaining $q$ Trotter steps can be neglected with minimal impact on the quality of the final solution. In Fig.~\ref{fig:hdcqc_ansatz}(a), we present a scheme of the purely quantum DCQO method, where the dominance of the CD term during the Trotterized evolution is highlighted based on the magnitude of $\alpha \dot{\lambda}$.

Furthermore, we implement a gate reduction strategy based on a user-defined threshold on gate angles. If the angle associated with a single or multi-qubit rotation is smaller than this threshold, the corresponding gate is not applied. This reduction is advantageous because the algorithm would only apply gates that are relevant enough to modify the state of the system up to a certain resolution given by the threshold. The reasonable threshold chosen for this study is a gate angle of less than $0.1$. Additionally, we note that for the IonQ hardware, the smallest gate angle their devices can accurately identify is $0.00628$~\cite{native_gates}. Setting such a threshold aligns well with the principle of operating within the impulse regime, elucidating why the omission of certain gates is effective in this scenario.

As a remark, one could further reduce the total gate counts by noting that the significance of the CD terms is mostly concentrated near the minimum gap, \( \Delta_{\text{min}} \). This region is where the majority of non-adiabatic transitions occur. Thus, it is essential to only apply the CD terms at these critical points, while determining the location of the minimum gap in advance poses a challenge. However, some recent proposals on adiabatic spectroscopy might be of help \cite{schiffer2022adiabatic}. We can therefore pinpoint those Trotter steps where the CD term dominates and eliminate the quantum gates where the CD term has minimal impact, ensuring efficiency without losing performance.
\subsection{Hybrid-DCQO}
Now, we propose the hybrid version of DCQO, which we call h-DCQO.
This algorithm differs from the previously introduced hybrid algorithm DC-QAOA \cite{chandarana2022digitized} in two aspects. 
First, a h-DCQO ansatz is composed of only the CD term~$\dot{\lambda} A_\lambda$, as opposed to the three terms in DC-QAOA~\eqref{eq:alternating_cd_qaoa}.
Second, in h-DCQO we consider one parameter for each CD term. A similar approach already showed interesting results on other use cases like protein folding ~\cite{chandarana2022protein}. 
Following the philosophy of variational quantum algorithms, the PQC generated from the h-DCQO ansatz is variationally optimized to minimize the expectation value.

As shown in Fig.~\ref{fig:hdcqc_ansatz}(b), a h-DCQO ansatz is parameterized by $\mathbb{\alpha}$ and $\mathbb{\beta}$, which correspond to the one- and two-body Hamiltonian terms, respectively, of the first order NC expansion of $A_\lambda$.  
Although both QAOA and h-DCQO ansatz are parameterized by the same number of parameters, the estimated two-qubit gate count for one layer of h-DCQO is double than that of QAOA, which is $pN(N-1)/2$.
In spite of this increased gate count per layer, h-DCQO outperforms QAOA by reducing the required number of layers for a target performance, as we demonstrate empirically in the next section.

\subsection{Model parameters and performance metric}
We use Markowitz's model for portfolio optimization in our experiments.
The coefficients of the problem Hamiltonian~\eqref{eq:hamiltonian} are calculated from the unconstrained formulation~\eqref{eq:portfolio}, with the Lagrange multipliers $\theta_1 = 1$, $\theta_2 = 0.5$ and $\theta_3 = 2$, 
and historical stock prices.
The values of the Lagrange multipliers are chosen by the user depending on how they emphasize the relative importance of the returns, risk and budget.
We obtained real data associated with 20 assets, labeled by \texttt{AAPL, JPM, JNJ, AMZN, PG, XOM, BA, DD, T, NEE, AMT, UPS, HD, PFE, NVDA, MSFT, GILD, GM, BRK-B} and \texttt{LMT}, for the period between 06.06.2022 and 01.01.2023 using the \texttt{YahooProvider} class from \texttt{Qiskit} framework~\cite{Qiskit}.
Additionally, we choose a reasonable value for the budget as $B= n/2$.

To quantify the performance of an algorithm for solving the portfolio optimization problem, we propose a `mean-based' definition of approximation ratio. Any quantum algorithm will give as output a state with energy $E$, where the optimal energy is $E_\text{min}$. It is common to use a definition of approximation ratio based on the spectral width ($E_\text{max}-E_\text{min}$) of the problem Hamiltonian~\cite{HSC+23}, where the assumed worst-case solution corresponds to the maximum eigen-energy~$E_\text{max}$. In practice, the worst case is rather obtaining the average energy $E_\text{avg}$ over a set of states where the solution is searched. 
Based on this, we modify the definition of approximation ratio to be 
\begin{equation}\label{mean_rat}
    r^\text{(avg)} \equiv \frac{E_\text{avg} - E}{E_\text{avg} - E_\text{min}} \, .
\end{equation}
We call this the mean-based approximation ratio and report our results using this metric. 
In particular, if the search set is an equal superposition of states~$\ket{+}^{\otimes n}$, i.e.\ randomly guessing any of the $2^n$ solutions with equal probability, $E_\text{avg}$ is zero and our expression reduces to $r^\text{(avg)} = E/E_\text{min}$. We can use different sets (e.g. set of feasible solutions), described by different values of $E_\text{avg}$, which establishes a reference for the worst case. 

\section{Implementation and Results}
\label{sec:results}
\begin{figure}
\includegraphics[width=0.9\columnwidth]{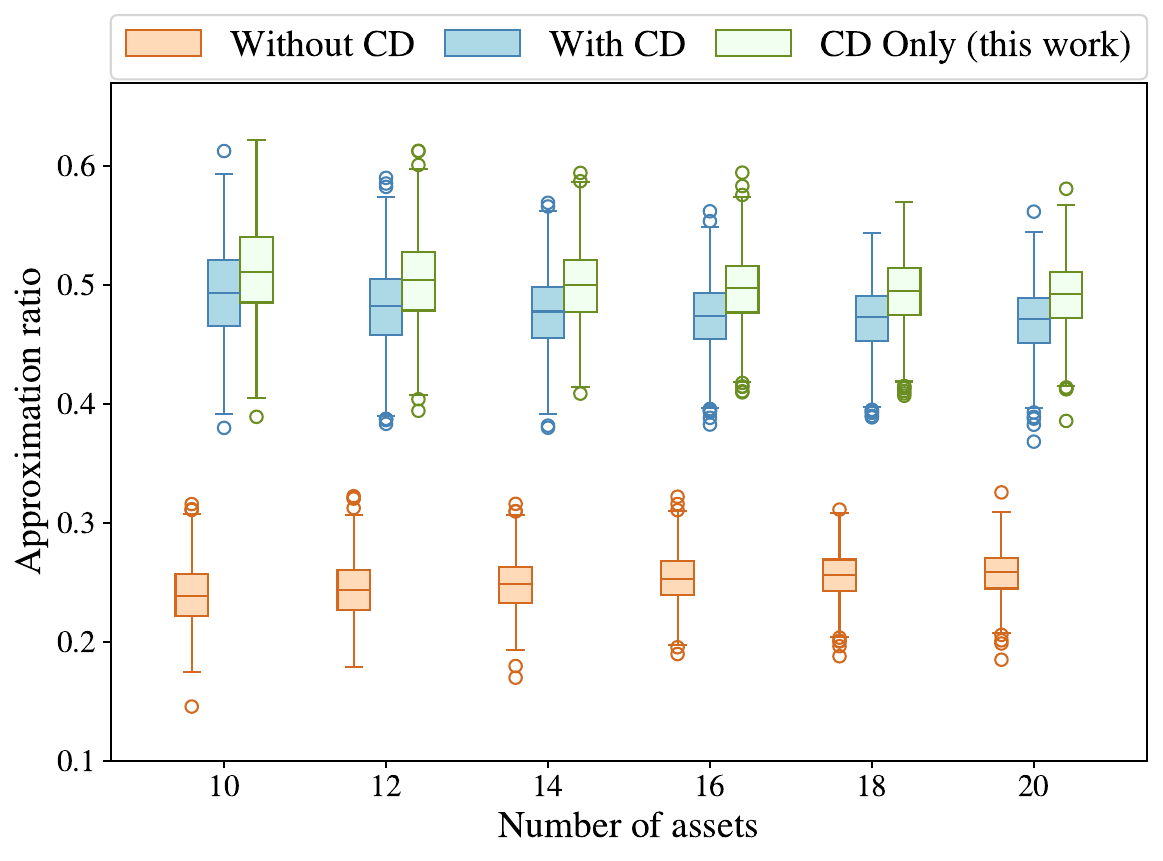}
  \caption{Comparison of the performance of three methods: non-adiabatic (\textit{Without CD}), inclusive of the first-order CD (\textit{With CD}), and the $1^{st}$ order CD term exclusively (\textit{CD only}). The data represents the approximation ratio for 1,000 randomly generated Ising spin-glass instances (refer to Eq. (\ref{eq:hamiltonian})) with 10 to 20 assets. The results were derived using a step size of $dt = 0.1$, with 12 Trotter steps in the \textit{without CD} case, 4 steps in the \textit{with CD} case, and 6 steps in the CD \textit{only} case, ensuring consistent circuit depth across all scenarios.}
  \label{fig:adbdcqo}
\end{figure}
We performed all our circuit implementations using BraKet SDK from AWS~\cite{braket}. 
For the hybrid algorithms, we leverage The Matter Lab’s Tequila framework~\cite{tequila} and implement an expectation value minimization routine using COBYLA as the optimization of choice and limiting to 200 iterations.
Given the all-to-all connectivity of the problem at hand, we select IonQ's trapped-ion device Aria, which has 25 fully connected qubits and is commercially available through AWS's BraKet service~\cite{braket}.
The average one-qubit and two-qubit gate errors are 0.06\% and 0.6\%, respectively~\cite{braket}.
\begin{figure*}[ht!]
  \includegraphics[width=0.9\linewidth]{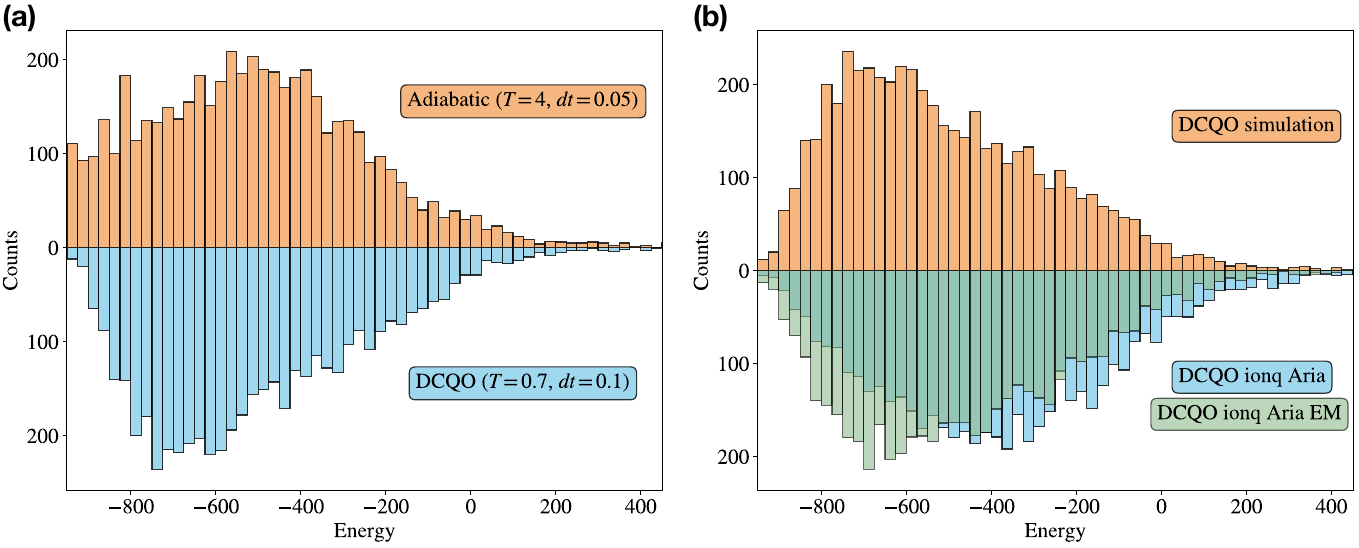}
  \caption{ (a) Energy distributions of the final state after applying DCQO (7 Trotter steps) and Adiabatic (80 Trotter steps). The approximation ratios are $0.54$ and $0.53$, respectively. (b) Energy distributions of the final state after applying DCQO in IonQ’s 25 qubits device Aria through AWS with and without error mitigation (debias) and the simulation results. The approximation ratios are $0.40$, $0.50$ and $0.54$, respectively. The number of shots was 5000 in both cases, which is the maximum supported by the hardware used. The gate cutoff threshold was set to 0.1}
  \label{fig:dcqc}
\end{figure*}

\subsection{Pure DCQO solutions}
In order to validate the DCQO (CD only) method, we examined 1000 random instances of the Ising spin-glass model with all-to-all interaction corresponding to the Hamiltonian in Eq.~\eqref{eq:hamiltonian}. The coefficients $h_i$ and $J_{ij}$ are drawn from a normal distribution with a mean of 0 and a variance of 1. We gathered results for systems of up to 20 qubits and compared the approximation ratios~\eqref{mean_rat} among the three purely quantum methods previously mentioned.
In Fig.~\ref{fig:adbdcqo}, we observe the advantage of using DCQO (CD Only) for fixed circuit depths. 
As expected, the addition of the CD term significantly improves the performance; in particular, a $2\times$ increase in the approximation ratio is observed.
The comparable performance of DCQO with and without the adiabatic Hamiltonian term also justifies the use of just the CD Hamiltonian in our further studies.

We present simulation results for a 20-asset portfolio optimization problem in Fig.~\ref{fig:dcqc}(a). In this figure, we observe that $7$ steps of DCQO, which reduce to $1$ step after gate cut-off, performs as good as $80$ steps of finite-time digitized-adiabatic evolution, since $r^\text{(avg)}\approx0.54$ for both cases. This means there is a $40\times$ reduction in the circuit depth as the depth-per-step for DCQO is twice that for finite-time digitized-adiabatic evolution.

For the hardware implementations, a set of modifications are performed to the logical representation in Fig.~\ref{fig:hdcqc_ansatz}. 
In particular, transpilation of logical gates to IonQ's native gates is performed using their native gate-set $\{GPi, GPi2, MS\}$; see Appendix \ref{appendix:a} for details. 
Additionally, the error mitigation~(EM) strategy of debiasing is added to reduce systematic errors on the hardware~\cite{maksymov2023enhancing}.
The performance gap between the simulated and hardware executions (without EM) is evident from Fig.~\ref{fig:dcqc}(b).
With EM, we are able to boost the performance to $r^\text{(avg)}=0.50$, which is comparable to the simulated result.
This highlights the fact that for sufficiently low-depth circuits, like that of DCQO (CD Only), debiased results from IonQ hardware are quite faithful to the ideal solutions.

\subsection{Hybrid DCQO solutions}
\begin{figure*}[ht!]
  \includegraphics[width=0.9 \linewidth]{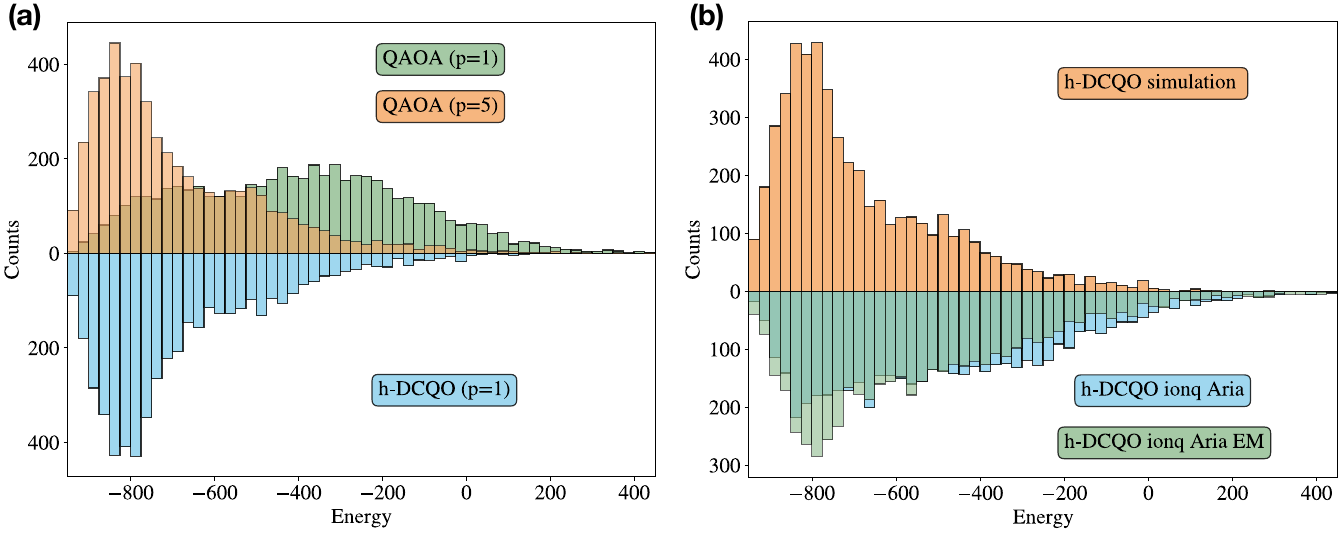}
  \caption{ (a) Energy distributions of the final state after applying h-DCQO (p = 1 layer) and QAOA (p = 1 and p = 5 layers) at the optimal parameters. The approximation ratios are $0.72$, $0.41$ and $0.73$, respectively. (b) Energy distributions of the final state after applying h-DCQO at the optimal parameters in IonQ’s 25 qubits device Aria through AWS with and without error mitigation (debias), and the simulation results. The approximation ratios are $0.58$, $0.52$ and $0.73$, respectively. Number of shots was 5000 in both cases, which is the maximum supported by the hardware used.}
  \label{fig:hdcqc}
\end{figure*}
In Fig.~\ref{fig:hdcqc}(a), we observe that a five-layer QAOA performs as good as a one-layer h-DCQO.
Although the depth per layer for h-DCQO is twice that of QAOA, this $5\times$ reduction in the required number of layers ultimately leads to a $2.5\times$ reduction in circuit depth for h-DCQO as compared to QAOA.
The reduction in the number of layers also leads to a five-fold reduction in the number of parameters, which, consequently, makes the variational training faster. 
A one-layer QAOA, on the other hand, has much worse performance.
Thus our h-DCQO algorithm is able to compress the required circuit depth for a target performance of $r^\text{(avg)}=0.72$ for the portfolio optimization problem.
The circuit-depth compression by h-DCQO is not only relevant for being able to implement any algorithm within the coherence time of the hardware but also to decrease the cumulative effect of hardware noise. 

In Fig.~\ref{fig:hdcqc}(b), we present results from our hardware implementation of h-DCQO and purposefully ignore QAOA due to its poor performance at low depth.
After performing the parameter learning locally, we execute the final circuit on IonQ's hardware.
While the mitigated results boost the section coinciding with the highest success probability states, the tail of the distribution is increased and noise mitigation does not seem to remove this, making the approximation ratio $r^\text{(avg)} = 0.58$. The existence of a significant gap between mitigated and unmitigated results, in the case of h-DCQO, is due to the fact that we do not apply the gate-cutoff threshold for h-DCQO, keeping the depth larger than DCQO circuits.
Thus, further noise suppression techniques are needed in order to close the gap between simulation and experiment (with EM) for the h-DCQO algorithm.

\section{Conclusions}
\label{sec:conclusion}
In this work, we demonstrated a drastic reduction in the quantum computational resources when tackling the portfolio optimization problem without compromising the solution quality. We achieved this important result by operating counterdiabatic dynamics in the fast-evolution or impulse regime, where the CD terms are dominant. Additionally, we outlined how to choose the corresponding Trotter steps so that the magnitude of the CD terms makes a significant contribution. Using the approximation ratio as a metric, we illustrated that both the pure quantum approach, DCQO, and the hybrid approach, h-DCQO, outperform digitized-adiabatic quantum protocols and QAOA. In this manner, we achieved a substantial reduction in the circuit complexity while maintaining a similar solution accuracy. Furthermore, we experimentally implemented our proposal on real-world portfolio data involving 20 assets and using a 20-qubit trapped-ion quantum computer. Finally, we validated the performance of our methods on a general Ising-spin glass problem with all-to-all interactions, underscoring their applicability to other combinatorial optimization problems.

As a final remark, we believe it is crucial to develop powerful methods to compress digital algorithms within the coherence time of available quantum computers. Furthermore, adopting a digital-analog approach may offer an intriguing new direction toward enhancing the practical utility of quantum computing in the noisy intermediate-scale quantum era.

\begin{acknowledgments}
We are grateful for the open access of Amazon Braket SDK and its associated service enabling access to IonQ's Aria device, as well as The Matter Lab's Tequila frameworks. The authors acknowledge Pranav Chandarana and Francisco Albarrán-Arriagada for useful discussions and feedback.
\end{acknowledgments}

\appendix
\section{Analytical calculation of CD coefficient from the first order nested commutator}
\label{appendix:a}
The first order CD coefficient can be computed exactly for any Hamiltonian, provided the first two nested commutators $\mathcal{O}_1 = [H_{ad}, \partial_\lambda H_{ad}]$ and $\mathcal{O}_2 = [H_{ad}, \mathcal{O}_1]$, i.e. $\alpha_1=-\Gamma_1 / \Gamma_2$, where $\Gamma_k=\left\|\mathcal{O}_k\right\|^2$ such that the CD term takes the form
\begin{equation}
    H_{C D}=i \dot{\lambda} \alpha_1 \mathcal{O}_1.
\end{equation}
For the case of the Hamiltonians used in Eq. (\ref{eq:adiabatic}),
\begin{equation}
    \begin{aligned}
\mathcal{O}_1 & =\left[-\sum_i X_i, \sum_k h_k Z_k+\sum_{k l} J_{k<l} Z_k Z_k\right] \\
& =2 i \sum_i h_i Y_i+2 i \sum_{i<j} J_{i j}\left(Y_i Z_j+Z_i Y_j\right).
\end{aligned}
\end{equation}
Therefore, $\Gamma_1$ is given by
\begin{equation}
   \Gamma_1=4 \sum_i h_i^2+4 \sum_{i<j} J_{i j}^2 \, .
\end{equation}
Now, we need to calculate the denominator of the equation for $\alpha_1$, which follows from  $\mathcal{O}_2=(1-\lambda)\left[H_i, \mathcal{O}_1\right]+\lambda\left[H_f, \mathcal{O}_1\right]$. Then, we need to evaluate two different commutators.
\begin{widetext}
The first one can be written as
\begin{equation}
    \left[H_i, \mathcal{O}_1\right]=4 \sum_i h_i Z_i+8 \sum_{i<j} J_{i j} Z_i Z_j-8 \sum_{i k j} J_{i j} Y_i Y_j \, ,
\end{equation}
whereas the second one is
\begin{equation}
    \begin{aligned}
{\left[H_f, \mathcal{O}_1\right] } & =4 \sum_i\left(h_i^2+\sum_j J_{i j}^2\right) X_i+8 \sum_{i \neq j} h_i J_{i j} X_i Z_j +8 \sum_{i<j<k}\left(J_{i j} J_{i j} X_i Z_j Z_k+J_{i j} J_{j k} Z_i X_j Z_k+J_{i k} J_{j k} Z_i Z_j X_k\right).
\end{aligned}
\end{equation}
It follows that,
\begin{equation}
\begin{aligned}
\Gamma_2= & 16(1-\lambda)^2\left(\sum_i h_i^2+4 \sum_{i \neq j} J_{i j}^2\right)+16 \lambda^2\left(\sum_i h_i^4+\sum_{i \neq j} J_{i j}^4+6 \sum_{i \neq j}^1 h_i^2 J_{i j}^2 +6 \sum_{i<j<k}\left(J_{i j}^2 J_{i k}^2+J_{i j}^2 J_{j k}^2+J_{i k}^2 J_{j k}^2\right)\right).
\end{aligned}
\end{equation}
Hence, the CD term is given by
\begin{equation}
    H_{C D}=-2 \dot{\lambda} \alpha_1\left(\sum_i h_i Y_i+\sum_{i<j} J_{i j}\left(Y_i Z_j+Z_i Y_j\right)\right), \; \; \; \text{where}\; \alpha_1=-\frac{1}{4} \frac{\sum_i h_i^2+\sum_{i<j} J_{i j}^2}{R(t)} ,
\end{equation}
and
\begin{equation}
\begin{aligned}
    R(t)=&(1-\lambda)^2\left(\sum_i h_i^2+4 \sum_{i \neq j} J_{i j}^2\right)+\lambda^2\left(\sum_i h_i^4+\sum_{i \neq j} J_{i j}^4+ 6 \sum_{i \neq j}^1 h_i^2 J_{i j}^2 +6 \sum_{i<j<k}\left(J_{i j}^2 J_{i k}^2+J_{i j}^2 J_{j k}^2+J_{i k}^2 J_{j k}^2\right)\right).
\end{aligned}
\end{equation}
\end{widetext}

\section{IonQ native gate implementation on Amazon Braket}
\label{appendix:b}
The transpilation process was entirely done finding optimal ways to translate the logical gate unitaries to their hardware-specific representation. In the case of the IonQ device, it is based on their reference material
\footnote{https://ionq.com/docs/getting-started-with-native-gates} 
where they represent their native $GPi, GPi2$, and $MS$ gate unitaries.

Logical to physical gates using single-qubit operations were translated looking for the appropriate parameter definition following the specification
\begin{widetext}
    \begin{equation*}
    GPi(\phi) = \begin{bmatrix}
    0 & e^{-i\phi}\\
    e^{i\phi} & 0
    \end{bmatrix}, \quad  \quad 
    GPi2(\phi) = \frac{1}{\sqrt{2}}\begin{bmatrix}
    1 & -ie^{-i\phi}\\
    -ie^{i\phi} & 1
    \end{bmatrix},
\end{equation*}
\begin{equation*}
    MS(\phi_0, \phi_1, \theta) = \begin{bmatrix}
    \cos\frac\theta2 & 0 & 0 & -i e^{-i(\phi_0 + \phi_1)} \sin\frac\theta2\\
    0 & \cos\frac\theta2 & -i e^{-i(\phi_0 - \phi_1)} \sin\frac\theta2& 0 \\
    0 &-i e^{i(\phi_0 - \phi_1)} \sin\frac\theta2 & \cos\frac\theta2 & 0 \\
    -i e^{i(\phi_0 + \phi_1)} \sin\frac\theta2  & 0 & 0 & \cos\frac\theta2
    \end{bmatrix}, 
\end{equation*}
\end{widetext}
which produces the replacements shown in Table~\ref{table1}.
\begin{table}
\caption{Transpilation of some important gates that are present in the Hamiltonians we worked with. Specifically, we implemented in Hardware the gates $H$, $R_y(\theta)$, $R_{yz}(\theta)$ and $R_{zy}(\theta)$ as shown in this table.}
\resizebox{\linewidth}{!}{%
\begin{tabular}{cc}
\hline
\hline
Logical gate & Native decomposition \\
 \hline
 $H$ & $GPi2(\pi/2) \times GPi(0)$ \\
 \hline
 $R_z(\theta)$ & $GPi(0) \times GPi(\theta/2)$ \\ 
 \hline
 $R_y(\theta)$ & $GPi2(\pi) \times GPi(\theta/2) \times GPi2(\pi)$ \\
 \hline
$R_{zz}(\theta)$ & $(GPi2(\pi) \otimes  GPi2(\pi))R_{yy}(\theta)(GPi2(0) \otimes  GPi2(0))$  \\ 
 \hline
$R_{zy}(\theta)$ & $(GPi2(\pi) \otimes  I)R_{yy}(\theta)(GPi2(0) \otimes  I)$ \\
 \hline
 $R_{yz}(\theta)$ & $(I \otimes  GPi2(\pi))R_{yy}(\theta)(I \otimes  GPi2(0))$  \\ 
 \hline
\end{tabular}
}
\label{table1}
\end{table}
Two qubit gates use the native M\o{}lmer-S\o{}renson gate as the basis of the translation. Depending on the angle of rotation different choices of the MS gate are selected based on their proximity to the extremes of a $2\pi$ circle. We used the definition of the partially entangling MS gate as described by IonQ, this requires three parameters in all cases where the third parameter must be $\theta \in [0, \frac{1}{4}]$ being the maximum value equal to the fully-entangled MS gate implementation. 
Table~\ref{table2} shows the implementation of the $R_{yy}(\theta)$ gate depending on the third parameter restriction on the angle being selected with respect to the $2\pi$ reference.
\begin{table}
\caption{Transpilation of the gate $R_{yy}(\theta)$ in terms of IonQ native gates for four different ranges of $\theta$.}
\resizebox{\linewidth}{!}{%
\begin{tabular}{cc}
\hline
\hline
Angle & Native gate \\
\hline
$\theta \leq \frac{\pi}{2}$ & $MS(\pi/2,\pi/2, \theta)$ \\ 
\hline
$\frac{\pi}{2} < \theta \leq \pi$ & $(GPi(\pi/2) \otimes  GPi(\pi/2))\times MS(3\pi/2,\pi/2, \pi-\theta)$ \\ [0.5ex] 
\hline
$\pi < \theta \leq \frac{3\pi}{4}$ & $(GPi(\pi/2) \otimes  GPi(\pi/2))\times MS(\pi/2,\pi/2, \theta-\pi)$ \\ [0.5ex] 
\hline
$\theta > \frac{3\pi}{2} $ & $MS(3\pi/2,\pi/2, 2\pi - \theta)$ \\ 
\hline
\end{tabular}
}
\label{table2}
\end{table}
Taking this as the basis, $R_{yy}(\theta)$ gate can be easily translated into different two-qubit gates needed for QAOA or hybrid DCQO by simply surrounding this action with the corresponding single gate rotations per qubit in the gate implementation.
Along with these translations, Amazon Braket SDK allows for debias for systematic error mitigation \cite{maksymov2023enhancing} by simply invoking it during the job submission. The main objective of this simple error mitigation technique was to not overcompensate by additional and more complex techniques.

\bibliography{reference.bib}
\end{document}